\documentclass[11pt, oneside]{article}   	
\usepackage{graphicx}
\usepackage{amssymb}
\usepackage{amstext}
\usepackage{mathrsfs}   
\usepackage{empheq}
\usepackage{epsfig}
\usepackage{tikz}
\usepackage{pdfpages}
\usepackage[section]{placeins}
\usepackage[pagebackref]{hyperref}

\def\x{\stackrel{\otimes}{,}}

\def\today{\number\day\space\ifcase\month\or Janvier \or F\'evrier \or  Mars 
   \or Avril \or Mai \or Juin \or Juillet \or Ao\^ut \or Septembre \or Octobre 
   \or Novembre \or D\'ecembre \fi\number \year}

 \makeatletter
 \@addtoreset{equation}{section}
 \makeatother

\newcommand{\f}[2]{{\ensuremath{%
    \mathchoice%
    {\dfrac{#1}{#2}}
    {\dfrac{#1}{#2}}
    {\frac{#1}{#2}}
    {\frac{#1}{#2}}
}}}

\newcommand{\tf}[2]{\ensuremath{#1/#2}}

\newcommand{\R}{\ensuremath{\mathbb{R}}}
\newcommand{\Cx}{\ensuremath{\mathbb{C}}}

\newcommand{\mc}[1]{\ensuremath{\mathcal{#1}}}
\newcommand{\mf}[1]{\ensuremath{\mathfrak{#1}}}
\newcommand{\msc}[1]{\ensuremath{\mathscr{#1}}}

\newcommand{\bs}[1]{\ensuremath{\boldsymbol{#1}}}

\newcommand{\sul}[2]{\ensuremath{\sum\limits_{#1}^{#2}}}
\newcommand{\pl}[2]{\ensuremath{\prod\limits_{#1}^{#2}}}

\newcommand{\op}[1]{ \boldsymbol{ \texttt{#1} } }
\newcommand{\wt}[1]{\ensuremath{\widetilde{#1}}}
\newcommand{\wh}[1]{\ensuremath{\widehat{#1}}}

\newcommand{\e}[1]{\ensuremath{\mathrm{#1}}}

\newcommand{\ex}[1]{\ensuremath{\e{e}^{#1}}}
\def \i{ \mathrm i}

\def\a{\alpha}

\def\be{\beta}
\def\ga{\gamma}

\def\de{\delta}

\def\De{\Delta}

\def\la{\lambda}

\def\sg{\sigma}

\def\th{\theta}

\def\Om{\Omega}
\def\om{\omega}
\def\vp{\varphi}

\newcommand\beq{\begin{equation}}
\newcommand\enq{\end{equation}}
\newcommand\bem{\begin{multline}}
\newcommand\enm{\end{multline}}

\def\ba{\begin{array}}
\def\ea{\end{array}}

\begin{document}

\title{The Toda$_2$ chain. }
\bigskip

\author{O. Babelon\footnote{Sorbonne Universit\'es, UPMC Univ Paris 06, CNRS, UMR 7589, LPTHE, F-75005 Paris, France}, K. K. Kozlowski\footnote{Univ Lyon, ENS de Lyon, Univ Claude Bernard Lyon 1, CNRS, Laboratoire de Physique,
 F-69342 Lyon, France}, V. Pasquier\footnote{Univ. Paris Saclay, CNRS, CEA, IPhT, F-91191 Gif-sur-Yvette, France.}}

\bigskip

\maketitle

\abstract{We show that a natural discretisation of Virasoro algebra yields a quantum integrable model which is the Toda chain in the second Hamiltonian structure. }

\section{Introduction}

It is well known that the classical Virasoro--Poisson bracket algebra is closely related to the so-called
exchange algebra \cite{Ba88,GN84}. The latter  admits a straightforward discretisation which allows one  to define, on the classical level, a discretised Virasoro-like theory \cite{Ba90}.
We observe that this theory coincides with the classical Toda chain in the second Hamiltonian structure (the Toda$_2$ chain), \textit{c.f.} \cite{ Adl79, Dam94}. 
As usual in the context of Toda lattice, the $N$-particle model can be formulated by means of a $N\times N$ single Lax matrix or through a product of
$2\times 2$ Lax matrices, both giving rise to the same spectral curve.

The quantisation of the exchange algebra is straightforward and provides a definition of the Toda$_2$ chain which we study
using the $ 2 \times 2$ Lax matrix formalism. The quantum Lax matrix $\op{l}_{0n}$ we obtain is not ultra-local. As a consequence, we need to resort to the Freidel-Maillet formalism \cite{FreMa91a,FreMa91b} so as 
to obtain the generating function of the quantum conserved quantities. 
This requires the introduction of supplementary scalar matrices $M_0$ \eqref{matrice de base scalaire solution FM eqns} and $\wt{M}_0$ \eqref{matrice de cloture de la trace scalaire solution FM eqns}
satisfying certain compatibility conditions. The most general solutions we find, exhibit a non-trivial dependence on $3$ free parameters. 
The Toda$_2$ chain arises as a specialisation of these parameters.

We choose an appropriate set of canonical operators to represent the Toda$_2$ algebra. Then, by means of 
a suitable gauge transformation, we show that we can relate the generating function of the quantum conserved quantities
to a new transfer matrix solely built in terms of ultralocalised Lax operators. This step brings us back to the standard formalism. 
We single out three special cases of the ultralocalised Lax matrix we obtained: the Toda$_2$ chain, the q-Toda chain and 
the q-oscillator model which coincides with the QTASEP stochastic model \cite{SW98}.

The paper is organised as follows. In Section \ref{Section Virasoro in Cont}, we recall two ways of constructing the classical Virasoro algebra 
in the continuum. In Section \ref{Section Lattice Virasoro}, we remind the construction of a natural discretisation of the classical Virasoro algebra. 
Then, in Section \ref{Section Toda2 classique}, we connect these results with the classical Toda chain endowed with the second Hamiltonian structure. 
In Section \ref{Section Toda2 quantique}, we construct the quantum counterpart of this model, the so-called Toda$_2$ chain. We show that although the resulting model is non-ultralocal, its  
quantum integrability can be described within the Friedel-Maillet scheme \cite{FreMa91a,FreMa91b}. 
Finally, in Section \ref{ultralocalisation}, we show that the Toda$_2$ chain transfer matrix ultralocalises, namely can be expressed in terms of an auxiliary transfer matrix 
associated with an ultralocal monodromy matrix.

\section{Classical Virasoro in the continuum}
\label{Section Virasoro in Cont}

Let $u$ be a $1$-periodic function on $\R$ that is endowed with the Virasoro Poisson bracket: 
\begin{eqnarray} 
\{u(x),u(y)\} = \f{1}{ 2} [u(x)+u(y)] \,  \delta'_{1}(x-y) \; + \;  \f{1}{2} \delta^{(3)}_{1}(x-y) \;,
\label{Vircont} 
\end{eqnarray} 
where $\de_{1}$ stands for the $1$-periodic Dirac Comb. Then, the re-scaled Fourier components of $u$
\beq
u(x) \, = \,  -48 \pi^2 \sul{ n \in \mathbb{Z} }{}  \big[ \mc{L}_n - \f{ \de_{n,0} }{ 24 } \big] \ex{2\i \pi n x}
\enq
satisfy the classical Virasoro Poisson algebra
\beq
48 \i \pi  \{ \mc{L}_n,\mc{L}_m \} \, = \, \mc{L}_{n+m} (n-m) \; + \; \f{n^3-n}{12} \de_{n+m,0} \;. 
\enq
The latter is obtained from the Virasoro algebra satisfied by the $\op{L}_n$'s by taking the $c\rightarrow \infty$  scaling  limit:
\[
\op{L}_n \rightarrow c \mc{L}_n \qquad \e{and} \qquad 
[\cdot, \cdot ] \rightarrow  48 \i \pi  c^{-1} \{\cdot, \cdot \}  \;. 
\]

Given $u$ endowed with the above bracket, consider the ordinary differential equation on $\R$
\begin{eqnarray} 
f^{\prime \prime}(x) + u(x) f(x) =0 \; . 
\label{Schroe} 
\end{eqnarray} 
It was shown in \cite{Ba88,GN84} that \eqref{Schroe} admits a basis of solutions $ \big\{ \xi^1(x),\xi^2(x) \big\}$, whose Wronskian is normalised to 1:
\beq
\det \left( \ba{cc}  \xi^1(x) &  \xi^2(x)   \vspace{2mm} \\ 
		   (\xi^1)^{\prime}(x) &  (\xi^2)^{\prime}(x)   \ea \right) \; = \; 1 \; ,  
\enq
and such that the vector valued row function $\xi(x)=(\xi^1(x), \xi^2(x) )$ has the Poisson bracket:
 \begin{eqnarray}  
\{\xi(x) \x \xi(y)\}=\xi(x) \otimes \xi(y) \cdot [ \mf{r}^+ \theta (x-y) +  \mf{r}^- \theta (y-x)] 
\label{Excont} 
\end{eqnarray} 
where  $\mf{r}^\pm$ are the $\mf{sl}_2$ solutions to the classical Yang-Baxter equation
\begin{eqnarray}
 \mf{r}^\pm = \pm  \big[ \sg^z\otimes  \sg^z + 4 \sg^{\pm} \otimes \sg^{\mp} \big] \;.  \nonumber
\end{eqnarray}
Here $\sg^z,\sg^{\pm}$ refer to the Pauli matrices
\begin{eqnarray}
\sg^z = \left( \ba{cc} 1 & 0 \\ 0 & -1 \ea \right) \; ,  \qquad \sg^+ =  \left( \ba{cc} 0 & 1 \\ 0 & 0 \ea \right) \;  ,  \qquad 
\sg^-=  \left( \ba{cc} 0 & 0 \\ 1 & 0 \ea \right)  \; .  \nonumber
\end{eqnarray}
Since $\xi^1, \xi^2$ solve \eqref{Schroe}, one can reconstruct  the potential $u(x)$ in terms of $\xi^1(x)$ and $\xi^2(x)$ by a Wronskian type relation 
\beq
u(x) \; = \; \det \left( \ba{cc}  (\xi^1)^{\prime}(x) &  (\xi^2)^{\prime}(x) \vspace{2mm} \\ 
		   (\xi^1)^{\prime\prime}(x) &  (\xi^2)^{\prime\prime}(x)   \ea \right)  \;. 
\label{equation u en terme des xis}
\enq
This observation allows one to recast \eqref{Schroe} in a determinantal form 
\begin{eqnarray} 
\det  \left( \ba{ccc} f(x) & \xi^1(x)  & \xi^2(x)   \\
	 		f^{\prime}(x)  &   (\xi^1)^{\prime}(x) &  (\xi^2)^{\prime}(x)  \\
			f^{\prime \prime}(x)   &    (\xi^1)^{\prime\prime}(x) &  (\xi^2)^{\prime\prime}(x)   \ea \right) =0 \nonumber 
\end{eqnarray}
In fact, one may turn things the other way around and consider the algebra \eqref{Excont} as a starting point. 
Then defining $u$   in terms of $\xi$ through \eqref{equation u en terme des xis}, one recovers the Virasoro--Poisson bracket eq.(\ref{Vircont}), see \cite{Ba88}.




\section{Classical Virasoro on the lattice}
\label{Section Lattice Virasoro}

The algebra (\ref{Excont}) corresponds to an equivalent formulation of the Virasoro--Poisson bracket. 
One of the advantages of such way of writing things is that this algebra can be naturally
put on the lattice, as it was shown in  \cite{Ba90}. We now recall this construction

Define row vector variables $\xi_n= (\xi^1_n,\xi^2_n)$ indexed by sites $n$ of a lattice and satisfying the  natural lattice version of exchange algebra eq.(\ref{Excont})
\begin{eqnarray} 
\{\xi_n \x \xi_m\}=\xi_n \otimes \xi_m \cdot  [ \mf{r}^+ \theta(n-m) + \mf{r}^- \theta (m-n)]
\label{Exlat}
\end{eqnarray}
where the Heavyside step function is defined as  $\theta(n)=1$ if $n>0$, $\theta(n)=0$ if $n<0$ and  $\theta(0)= \tf{1}{2}$. 

The associated lattice Wronskians
\begin{eqnarray}
W^{(p)}_n=\xi^1_n\; \xi^2_{n+p}-\xi^2_n\; \xi^1_{n+p}   \nonumber 
\end{eqnarray}
form a closed algebra under the Poisson bracket only if $p=1$ and $2$. The latter takes the form 
\begin{eqnarray}
\{ W^{(1)}_n, W^{(1)}_m\}&=&\; W^{(1)}_n\;
W^{(1)}_m\;(\delta_{n,m-1}-\delta_{n,m+1})   \; ,    \label{w1w1} \\ 
\{ W^{(1)}_n,W^{(2)}_m\}&=& \;W^{(1)}_n\;
W^{(2)}_m\;(\delta_{n,m+1}-\delta_{n,m+2}+\delta_{n,m-1}-\delta_{nm}) \; , 
\label{w1w2} \\ 
\{ W^{(2)}_n, W^{(2)}_m\}&=&\;  W^{(2)}_n \;W^{(2)}_m\;(\delta_{n,m-2}
-\delta_{n,m+2}+2\delta_{n,m+1}-2\delta_{n,m-1}) \nonumber \\
& &-4\; W^{(1)}_{n-1}\; W^{(1)}_{n+1}\;\delta_{n,m+1}
        +4\; W^{(1)}_{m-1}\; W^{(1)}_{m+1}\;\delta_{n,m-1} \; . 
 \label{w2w2}
\end{eqnarray}
Upon introducing 
\begin{eqnarray}
S_n \, = \, 4 \f{ W^{(1)}_{n+1} W^{(1)}_{n-1} }{ W^{(2)}_{n} W^{(2)}_{n-1} } 
\nonumber  
\end{eqnarray}
the above  algebra separates into
\begin{eqnarray}
\{ W^{(1)}_n, W^{(1)}_m\}&=& W^{(1)}_n\;
W^{(1)}_m\;(\delta_{n,m-1}-\delta_{n,m+1})   \label{u1} \\ 
\{ W^{(1)}_n, S_m\}&=& 0  \nonumber   \\ 
\{ S_n, S_m\}&=&-
S_nS_m[(4-S_n-S_m)(\delta_{n,m-1}-\delta_{n,m+1}) \nonumber \\
   & &\qquad \qquad \quad +S_{n-1}\delta_{n,m+2}-S_{m-1}\delta_{n,m-2} ]     \;. 
\label{Virlat} 
\end{eqnarray}
This separability has nonetheless a  price in that the resulting subalgebra  $\{S_n \}_{n \in \mathbb{Z} }$ is cubic as opposed to the quadratic nature of the  $W$-algebra.  
The $S_n$ algebra was first obtained in \cite{FaTa86,V92}. It may be considered as a lattice deformation of the Virasoro algebra. Indeed, setting $S_n \, = \, 1+ \tfrac{ \De^2 }{ 2 }   u_{\De}(n\De)$, 
and imposing the scaling in the $\De \rightarrow 0^+$ limit, 
\beq
 u_{\De}(n\De) \underset{ n \De \rightarrow x }{ \longrightarrow } u(x) \qquad \e{and} \qquad  \{ \cdot , \cdot \} \rightarrow  16 \{ \cdot , \cdot \}_0
\enq
where $ \{ \cdot , \cdot \}_0$ denotes the bracket in the continuum that appeared in Section \ref{Section Virasoro in Cont}, one gets that \eqref{Virlat} goes to \eqref{Vircont}.

\bigskip

\section{The classical Toda$_2$ chain}
\label{Section Toda2 classique}

By analogy with the continuum case, one may interpret the quantities $\xi_n^1$ and $\xi_n^2$ as two independent solutions of a three term linear
recursion  which can be put in the form 
\begin{eqnarray} 
\det  \left( \ba{ccc}   v_n      & \xi_n^1      &  \xi_n^2      \\
             v_{n+1}  & \xi_{n+1}^1  &  \xi_{n+1}^2  \\
              v_{n+2}  & \xi_{n+2}^1  &  \xi_{n+2}^2  \ea \right) \, = \, 0 \;, 
\nonumber 
\end{eqnarray}
or, equivalently, 
\begin{eqnarray}
W^{(1)}_n v_{n+2} -W^{(2)}_n v_{n+1} +W^{(1)}_{n+1} v_n \, = \, 0 \;. 
\label{SchW}
\end{eqnarray}
It appears convenient to recast eq.(\ref{SchW}) in terms of new quantities 
\begin{equation}
Q_n= {1\over W^{(1)}_{n}}, \quad P_{n}={W^{(2)}_{n-1}\over W^{(1)}_{n-1}W^{(1)}_{n}} \, , 
\label{QPclas}
\end{equation}
which have simpler Poisson brackets than the $W_n^{(a)}$'s:  
\begin{eqnarray}
\{Q_n, Q_m\} &=& Q_n Q_m ( \delta_{n+1,m} -\delta_{n,m+1} ) \; , \label{QQ}\\
\{Q_n, P_m\} &=& -2 Q_n P_m ( \delta_{n,m} - \delta_{n+1,m} )  \; , \label{QP} \\
\{P_n, P_m\} &=& -4 Q_{m}^2 \delta_{n,m+1} +4 Q_{n}^2 \delta_{n+1,m} \label{PP} \;. 
\end{eqnarray}
Then,  eq.(\ref{SchW}) takes the form
\begin{eqnarray}
Q_{n}v_{n+1} - P_n v_{n} + Q_{n-1}v_{n-1} =0 \;. 
\label{SchWbis}
\end{eqnarray}

When imposing periodic boundary conditions for the $P_n$'s and $Q_{n}$'s: $P_{n+N}=P_n$ and $Q_{n+N}=Q_n$ and quasi-periodic ones for the $v's$ $v_{N+n}=\mu v_n$,
the linear system eq.(\ref{SchWbis}) can be put in an $N\times N$ matrix form:
\begin{empheq}{equation}
L(\mu) \cdot \vec{v}= \left( \ba{cccccc}  -P_1 & Q_1 & 0 & \cdots & 0 &\mu^{-1} Q_N \\
            Q_1 & -P_2 & Q_2 & 0 & \cdots &0  \\ 
             0 & Q_2 & -P_3 & Q_3 & 0 & \vdots \\
             \vdots   &    0   &  \ddots     &    \ddots    &   \ddots   &     0     \\
             0   &                   &   0   &       &   & Q_{N-1 }           \\
               \mu Q_N  &0  &\cdots   &     0 & Q_{N-1 }      &  -P_N    \ea \right) \cdot            
\left( \ba{c}  v_1 \\ v_2 \\  v_3 \\ \vdots      \\ v_N \ea \right)  \,  = \, 0       \;.    
\label{bigLax}
\end{empheq}

Note that under periodic boundary conditions for the $Q_n$, $P_n$'s, one has to understand the Kronecker symbols in \eqref{QQ}-\eqref{PP} modulo $N$, \textit{viz}. $\de_{n,m}=1$ if and only if $n-m \in N \mathbb{Z}$.

The above Jacobi matrix is typical of the Lax matrix of the Toda chain. Furthermore, the quantities  ${\rm tr} \big[ L^n(\mu) \big]$ are  in involution. 
 Indeed, one can check that the Poisson bracket of $L$ is of the form
$$
\big\{ L_1(\mu_1) \x  L_2(\mu_2)  \big\}  \, = \,  \big[  \mf{d}_{12},L_1(\mu_1) \big] \, - \, \big[ \mf{d}_{21} , L_2(\mu_2) \big] \qquad \e{with} \quad
\left\{ \ba{ccc} L_1(\mu) &=&\op{id}\otimes L(\mu)    \vspace{2mm} \\  L_2(\mu) & = & L(\mu)\otimes  \op{id} \ea  \right.
$$
ensuring the Poisson commutativity in question. We find
$$
\mf{d}_{12} = - \big( \mf{r}_{12} - \mf{a}_{12}\big) \cdot  L_2(\mu_2)  \, - \,  L_2(\mu_2) \cdot \big( \mf{r}_{12} + \mf{a}_{12} \big)
$$ 
where, using the elementary $N\times N$ matrices $\op{E}_{ij}$, one has  
\begin{eqnarray}
\mf{r}_{12}  & = & \sul{j>i}{N} \Big[  \f{2\mu_2}{\mu_1-\mu_2} \op{E}_{ij}\otimes\op{E}_{ji}  + \f{2\mu_1}{\mu_1-\mu_2} \op{E}_{ji}\otimes\op{E}_{ij} \Big] 
\; + \;  \sul{i=1}{N}  \f{\mu_1+\mu_2}{\mu_1-\mu_2} \op{E}_{ii}\otimes\op{E}_{ii}  \; ,  \\
\mf{a}_{12}  & = & \f{1}{2} \sul{j>i}{N} \Big[   \op{E}_{ii}\otimes\op{E}_{jj}  - \op{E}_{jj}\otimes\op{E}_{ii} \Big] \; . 
\end{eqnarray}
 Explicitly, we have
\begin{empheq}{align}
\big\{ L_1(\mu_1) \x L_2(\mu_2) \big\} \, =&~~~ \, \big[ -2 \mf{r}_{12} , L_1 (\mu_1)L_2(\mu_2) \big] \nonumber \\
& +( 2 \mf{a}_{12} ) L_1(\mu_1) L_2(\mu_2)
+   L_1(\mu_1) L_2(\mu_2) (2 \mf{a}_{12}) \nonumber \\
& - 2  L_1(\mu_1)\,  \mf{a}_{12} \,  L_2 (\mu_2)
-  2  L_2(\mu_2) \,  \mf{a}_{12} \, L_1(\mu_1) 
\label{poissonL}
\end{empheq}
Hence, the finite difference equation \eqref{SchWbis} is an integrable system belonging to the Toda family. 
In fact, the bracket given in eqns \eqref{QQ}-\eqref{PP}  is the well known  second Poisson bracket of the Toda chain \cite{Adl79,Dam94}. 

The spectral curve of the model is then defined as 
\beq
0 \, =  \, \e{det}_N\big[ L(\mu) +\la \, \e{id} \big] \, = (-1)^{N+1}\, \pl{a=1}{N}  Q_{a} \cdot \big\{  \mu + \mu^{-1}  \big\} \, + \, p_N(\la)
\enq
where $p_N$ is a monoic polynomial in $\la$ of degree $N$. 

\vspace{2mm}

Alternatively, one can embrace the classical integrability of the model by means of the $2\times 2$ Lax matrix
\begin{equation}
l_n(\lambda) = \left( \ba{cc}  \lambda - P_n & -1 \\ Q_n^2 &  0 \ea \right) \;. 
\label{Laxmatrixln}
\end{equation}
Since we will provide a thorough discussion of the integrability in the quantum case, we leave aside the details of the construction of the $2\times 2$
monodromy matrix and the one of the generating function of conserved quantities.  The bottom line is that, in such a case, the monodromy matrix takes the form 
\beq
T_N(\lambda)  \, = \,  l_N(\lambda) l_{N-1}(\lambda) \cdots l_1(\lambda) \;. 
\enq
The associated spectral curve takes the form 
\beq
0\, = \, \mu^{-1} \cdot \e{det}_{2}\Big[   T_N(\la)  \,  - \, \mu \,  \e{id}   \Big] \, = \, \mu \,  + \,  \pl{a=1}{N}  Q_{a}^{2}  \cdot  \mu^{-1}  \, - \,    \e{tr}\big[ T_N(\la) \big] \; . 
\enq
An explicit calculation shows that it should hold $p_N(\la) \, =\,  \e{tr}\big[ T_N(\la) \big]$. 
Thus, upon noticing that $\prod_{a=1}^{N}Q_a$ is a conserved quantity and hence plays the role of a scalar quantity, one may rescale $\mu \hookrightarrow \mu (-1)^{N}  \prod_{a=1}^{N}  Q_{a} $
in the monodromy matrix based spectral curve so as to get a full identification of the two spectral curves introduced above.

\vspace{3mm}

 One can represent $\xi_n^{(1,2)}$ in terms of canonical Darboux coordinates $\{x_n,X_m\}= 4 \delta_{nm}$ :
\begin{equation}
\xi^1_n \, = \,  \ex{-{1 \over 2}x_n }  \pl{a=1}{n} \ex{ {1\over 2} X_a},\quad 
\xi^2_n = \, \ex{ -{1 \over 2}x_n }\sum_{a=1}^n \pl{b=a}n \ex{  {1 \over 2} X_b}  \cdot  \ex{x_a}  \cdot \pl{b=1}{a-1} \ex{-  {1 \over 2} X_b}
\label{DarbouxCoordinates}
\end{equation}
In terms of these canonical coordinates, one has
\begin{eqnarray}
Q_n^2 &=& \ex{-X_{n+1} } \ex{x_n -x_{n+1}} \; , \label{RepQn} \\
P_n &=&  e^{-X_n } + \ex{x_n -x_{n+1}} \; . \label{RepPn}
\end{eqnarray}
Within this parametrisation  and upon setting 
\beq
v_n \, = \, f_{\De}\big(  n \De   \big) \qquad X_n=2 \De \pi_{\De}(n\De) \qquad x_n=\vp_{\De}(n\De) 
\enq
consider the continuum limit 
\beq
N\De \rightarrow 1 \qquad \e{and} \qquad 
\left\{ \ba{ccc} \pi_{\De}(n\De) &\rightarrow & \pi(x) \vspace{2mm} \\
		\vp_{\De}(n\De) &\rightarrow & \vp(x) \ea \right. 
\qquad \e{so} \, \e{ that}  \quad \{\vp(x),\pi(x)\}=2 \de(x-y) \, .  
\enq
Then,  eq.(\ref{SchWbis}) yields
\begin{equation}
f^{\prime\prime}(x) \, - \, \left[\left(\pi(x)-{1\over 2} \varphi^{\prime}(x) \right)^2 +  \left(\pi(x)-{1\over 2} \varphi^{\prime}(x)\right)^{\prime} \right] f(x) \, = \, 0 \;. 
\label{schroecont}
\end{equation}
Hence, $p(x)=\pi(x)-{1\over 2} \varphi^{\prime}(x)$ is the standard chiral Coulomb field in CFT and eqs.(\ref{DarbouxCoordinates}) are lattice analogues of the standard 
representation of the solutions of eq.(\ref{schroecont}) by classical vertex operator and screening.

 \bigskip

\section{ The quantum Toda$_2$ chain}
\label{Section Toda2 quantique}

In the quantum case the exchange algebra for  the  row vectors $\bs{\xi}_n=(\bs{\xi}_n^1,\bs{\xi}_n^2 )$ reads
\beq
\bs{\xi}_{n} \otimes  \bs{\xi}_{m} = \bs{\xi}_m \otimes \bs{\xi}_n \cdot  \op{P}\,  [ \, \op{R}^+ (q) \, \theta_q(n-m) \, + \,  \op{R}^-(q) \, \theta_q(m-n) ]
\label{ecriture algebre des xi}
\enq
where $\op{P}$ is the permutation matrix on $\Cx^2\otimes \Cx^2$, 
\begin{eqnarray}
\op{R}^+(q) = \left( \ba{cccc} q^{{1\over 2}}  &    0     &     0       &  0  \\
                0     &  q^{-{1\over 2}}    & q^{{1\over 2}}-q^{-{3\over 2}}   &  0  \\
                0     &    0       &    q^{-{1\over 2}}   &  0  \\
                0               &    0       &     0       &  q^{{1\over 2}}  \ea \right) \quad  \e{and} \quad \op{R}^-(q)=\op{P} \, \op{R}^+(q^{-1})\, \op{P} \;. 
\nonumber
\end{eqnarray}
Finally, $\th_q$ is the $q$-deformation of the Heaviside function:
\beq
\th_q(x)=0 \quad \e{if}\quad x<0 \; , \qquad \th_q(x)=1 \quad \e{if}\quad x>0 \qquad \e{and} \qquad  \th_q(0)= \f{ 1 }{ q^{ \frac{1}{2} } + q^{ -\frac{1}{2} }  } \;  .  
\enq

Since the $\op{R}$-matrix is independent of the lattice spacing $\Delta$, it can be computed in the continuous CFT theory \cite{GN84}. The parameter $q$ is related to the central charge by :
$$
c=1+6 \left( b+{1\over b}\right)^2, \quad q= \ex{ \i \pi b^2},\quad \tilde{q}= \ex{{ \i \pi \over b^{2}}} \;. 
$$
They form a dual pair in the sense of Faddeev's modular double theory. The range of $b$ is the real axis $b>1$ for $c > 25$, the unit circle 
$0 < {\rm Arg}\; b <  \pi/2$ for  $1 < c < 25$ and the imaginary axis ${\rm Im}\; b >1$  for $c < 1$. In the following, it appears convenient to choose the below parametrisation of $b$
\beq
b^2 \, = \, \f{\om_1}{\om_2} \;. 
\enq
Also, in our further considerations, we will consider any value of $\om_1, \om_2$, \textit{viz}. $b$,  without limiting ourselves to the values giving rise to a natural CFT interpretation.

\bigskip

We  define as in \cite{Ba90} the quantum $W$-algebra. Let
$$
\op{W}_n^{(p)} =  q\;  \bs{\xi}^1_n \bs{\xi}^2_{n+p} \, -\,  \bs{\xi}^2_n \bs{\xi}^1_{n+p}, \quad p=1,2 \;. 
$$
Then the  quantum $W$-algebra reads
\begin{eqnarray}
\op{W}^{(1)}_n\; \op{W}^{(1)}_m &=& q^{{1\over 2}(\delta_{n,m-1}-\delta_{n,m+1})}\;\op{W}^{(1)}_m\; \op{W}^{(1)}_n \; , \nonumber \\
\op{W}^{(1)}_n\; \op{W}^{(2)}_m &=& q^{{1\over 2}(-\delta_{n,m+2}+ \delta_{n,m+1}-\delta_{nm}+\delta_{n,m-1})}\;\op{W}^{(2)}_m \;\op{W}^{(1)}_n \; , 
\nonumber \\
\op{W}^{(2)}_n\; \op{W}^{(2)}_m &=& q^{{1\over 2}(\delta_{n,m-2}-\delta_{n,m+2}+2\delta_{n,m+1}- 2\delta_{n,m-1})}\;\op{W}^{(2)}_m\; \op{W}^{(2)}_n
\nonumber \\
& & +(q^{-{1\over 2}}-q^{{3\over 2}})\;\op{W}^{(1)}_{n-1} \;\op{W}^{(1)}_{n+1}\;
\delta_{n,m+1}
     +(q^{{1\over 2}}-q^{-{3\over 2}})\;\op{W}^{(1)}_{m-1} \;\op{W}^{(1)}_{m+1}\; \delta_{n,m-1} \; . \nonumber     
\end{eqnarray}
The quantum analogues  $\op{P}_n,\op{Q}_n$ of the classical variables given in eq.(\ref{QPclas}) are constructed as 
$$
\op{Q}_n  \, = \,  [\op{W}_n^{(1)}]^{-1}, \quad \op{P}_n = \op{Q}_{n-1} \op{Q}_n  \op{W}^{(2)}_{n-1} = \op{W}^{(2)}_{n-1} \op{Q}_{n-1} \op{Q}_n \; , 
$$
and enjoy the commutation relations
\begin{eqnarray*}
\op{Q}_n \op{Q}_m &=& q^{{1\over 2} (\delta_{n,m-1} - \delta_{n,m+1})} \op{Q}_m \op{Q}_n  \; , \\
\op{P}_n \op{P}_m &=&  \op{P}_m \op{P}_n + (q^{3\over 2} - q^{-{1\over 2}} ) ( \op{Q}_n^2 \delta_{n+1,m} \, -\,  \op{Q}_m^2 \delta_{n,m+1} )  \; , \\
\op{P}_n \op{Q}_m &=&  q^{(\delta_{n,m} - \delta_{n,m+1})} \op{Q}_m \op{P}_n \; .  
\end{eqnarray*}
The quantum Lax matrix takes a   form analogous to the classical case
\begin{equation}
\op{l}_{0n}(\lambda) \, = \,  \left( \ba{cc}  \lambda - \op{P}_n & -1 \\ \op{Q}_n^2 &  0 \ea \right)_{[0]}
\label{Laxmatrixquantum}
\end{equation}
where $0$ indexes the auxiliary space. Since the model is non-ultralocal, we must use the Freidel-Maillet  scheme \cite{FreMa91a,FreMa91b}
to  encode, in an integrable fashion, the commutation relations for the quantum Lax matrix:
\begin{eqnarray}
{\cal A}_{12}(\lambda_1,\lambda_2) \, \op{l}_{1n}(\lambda_1) \, \op{l}_{2n}(\lambda_2)&=& \op{l}_{2n}(\lambda_2)  \, \op{l}_{1n}(\lambda_1) \, {\cal D}_{12}(\lambda_1,\lambda_2)  \; , 
\label{AD}  \vspace{3mm} \\
\op{l}_{1n}(\lambda_1)\, \op{l}_{2,n+1}(\lambda_2)&=&\op{l}_{2,n+1}(\lambda_2) \, {\cal C}_{12}(\lambda_1,\lambda_2)  \, \op{l}_{1n}(\lambda_1)  \; , 
 \label{B} \vspace{3mm} \\
\op{l}_{2n}(\lambda_2) \, \op{l}_{1,n+1}(\lambda_1) &=& \op{l}_{1,n+1}(\lambda_1) \,   {\cal B}_{12}(\lambda_1,\lambda_2) \,  \op{l}_{2n}(\lambda_2) \; . 
 \label{C}
\end{eqnarray}
The matrices ${\cal A}_{12}(\lambda_1,\lambda_2)$ ${\cal B}_{12}(\lambda_1,\lambda_2)$, ${\cal C}_{12}(\lambda_1,\lambda_2)$ and ${\cal D}_{12}(\lambda_1,\lambda_2)$ 
take the form
$$
{\cal A}_{12}(\lambda_1,\lambda_2)=\left( \ba{cccc} 1 & 0 & 0 & 0 \\
	  0&{{\left(\lambda_2 -\lambda_1 \right)\,}\over{\lambda_2 \,q^2-\lambda_1 }}&{{\lambda_1 \left(q^2-1\right)}\over{\lambda_2 \,q^2-\lambda_1  }}& 0 \\
	0&{{\lambda_2 \,\left(q^2-1\right) }\over{\lambda_2 \,q^2-\lambda_1 }}&{{\left(\lambda_2 -\lambda_1  \right)\,q^2}\over{\lambda_2 \,q^2-\lambda_1 }} &  0  \\
	  0  &  0  &  0  &   1   \ea \right) \;, 
$$
 $$
{\cal  D}_{12}(\lambda_1,\lambda_2)   =   \left( \ba{cccc}   1&0&0&0\\
    0&{{\left(\lambda_2 -\lambda_1 \right) q^2}\over{\lambda_2 \,q^2-\lambda_1 }}&{{\lambda_2   \,\left(q^2-1\right)}\over{\lambda_2 \,q^2- \lambda_1 }} & 0  \\
      0&{{\lambda_1 \,\left(q^2-1\right)}\over{\lambda_2 \,q^2-\lambda_1 }}&{{\left(\lambda_2 -\lambda_1 \right)}\over{\lambda_2 \,q^2-\lambda_1 }}&0 \\ 
0 &{{\lambda_1 \,\left(\lambda_2 -\lambda_1 \right)\,\left(q^2-1\right)}\over{\lambda_2 \,q^2-\lambda_1 }}&-{{ \lambda_2 \,\left(\lambda_2 -\lambda_1 \right)\,\left(q^2-1 \right)}\over{\lambda_2 \,q^2-\lambda_1 }}&1
 \ea \right) \;, 
 $$
 $$
{\cal  C}_{12}(\lambda_1,\lambda_2)  =   \left( \ba{cccc}  1 & 0 &  0 & 0   \\
		0&1& -\left(q^{3/2}-q^{-1/2}\right)&0    \\
					    0&0&q^2&0  \\
	      0&0& \lambda_2\, \left(q^2-1\right)&1   \ea \right) \;, 
 $$

 $$
{\cal  B}_{12}(\lambda_1,\lambda_2)  =   \left( \ba{cccc}   1 & 0 & 0 & 0   \\
						      0 &q^2 & 0 & 0  \\
					0 & -\left(q^{3/2}-q^{-1/2}\right)\, & 1 & 0 \\ 
					0 & \lambda_1\,\left(q^2 -1\right) & 0 & 1 \ea \right) \;. 
 $$

Following \cite{FreMa91a, FreMa91b}, the construction of the model's monodromy matrix demands to find a  numerical matrix  $M_0(\lambda)$  on the auxiliary space obeying to the compatibility relations
\begin{equation}
{\cal D}_{12}(\la_1,\la_2) M_1(\la_1) {\cal C}_{12}(\la_1,\la_2) M_2(\la_2) = M_2(\la_2) {\cal B}_{12}(\la_1,\la_2) M_1(\la_1) {\cal A}_{12}(\la_1,\la_2) \;. 
 \label{DGCG}
\end{equation}
The most general solution to \eqref{DGCG} takes the form 
\beq
M_0(\la) \; = \; \a \left(\ba{cc} 1  &  \be \la \\ 
				\ga \la & q^{-\frac{1}{2}}+\de \la +\be  \la^2 \ea \right)_{[0]} \;, 
\label{matrice de base scalaire solution FM eqns}
\enq
for arbitrary constants, $\a,\be,\ga,\de$. 
Then, the monodromy matrix is defined as 
\begin{equation}
\op{l}_{0N}(\lambda)\, M_0(\la)\,  \op{l}_{0 N-1}(\lambda)\, M_0(\la) \,  \cdots \, \op{l}_{02}(\lambda)\, M_0(\la)\, \op{l}_{01}(\lambda)
\end{equation}
where the matrix $  \op{l}_{0n}(\lambda) M_0(\la)$ takes the explicit form
\begin{empheq}{equation*}
\wh{\op{l}}_{0n}(\lambda)  =  \alpha
\begin{pmatrix} (1-\gamma) \lambda -\op{P}_n & -q^{-1/2} -\delta \lambda -\lambda \beta \op{P}_n \cr
 \op{Q}_n^2 & \lambda \beta \op{Q}_n^2 \end{pmatrix} \;. 
\end{empheq}
We see that the constant $\gamma$ can be reabsorbed into a redefinition of $\lambda, \beta, \delta$
and plays no role (the case $\gamma=1$ should be studied separately). The classical $l_n(\lambda)$, eq.(\ref{Laxmatrixln}), is recovered for $\beta=\delta=0$ (and $q=1$).
The solution of interest to our analysis is conveniently parameterised  by taking  
\beq
\a=1\;,  \quad  \be =q^{\frac{7}{2}} d_{2} d_{3} \; , \quad \ga = (1-q^{2}) \; ,  \quad  \de = q^{\frac{5}{2}} d_{1}  
\enq
where  $d_{1}, d_{2}, d_{3}$ are arbitrary constants. We denote 
this specific solution of \eqref{DGCG} as
\begin{equation}
G_0(\lambda) =  \left( \ba{cc} 1 & q^{ \frac{7}{2} } d_{2} d_{3} \la    \vspace{3mm} \\ (1-q^2)\lambda &  q^{- \frac{1}{2} } + q^{ \frac{5}{2} } d_{1} \lambda + q^{ \frac{7}{2} } d_{2} d_{3} \la^2 \ea \right) \;. 
 \label{gammamatrix}
\end{equation}
Thus, we shall focus below on the monodromy matrix  
\begin{equation}
\op{T}_{0N}(\lambda) \, = \,  \wh{\op{l}}_{0N}(\lambda)\,  \wh{\op{l}}_{0 N-1}(\lambda)\, \cdots \, \wh{\op{l}}_{02}(\lambda)\, \ \op{l}_{01}(\lambda)
 \label{transfer}
\end{equation}
where $ \wh{\op{l}}_{0n}(\lambda)=  \op{l}_{0n}(\lambda) G_0(\la)$.

The exchange algebra \eqref{AD}-\eqref{C} and the equation \eqref{DGCG} put together ensure that the above monodromy matrix satisfies the quadratic algebra 
\begin{equation}
{\cal A}_{12}(\la_1,\la_2) \op{T}_{1N}(\la_1) {\cal B}_{12}(\la_1,\la_2)   \op{T}_{2N}(\la_2) \; = \;   \op{T}_{2N}(\la_2) {\cal C}_{12}(\la_1,\la_2)   \op{T}_{1N}(\la_1) {\cal D}_{12}(\la_1,\la_2) \;. 
 \label{ATT=TTD}
\end{equation}

The associated family of commuting quantities can be constructed \cite{FreMa91a, FreMa91b} by using a  numerical matrix  $ \wt{M}_0(\lambda)$  on the auxiliary space obeying to the 
dual compatibility relation:
\beq
{\cal D}_{12}(\la_1,\la_2) \wt{M}_2(\la_2) \, \wt{\cal B}_{12}(\la_1,\la_2) \, \wt{M}_1(\la_1) \; = \;  \wt{M}_1(\la_1) \wt{\cal C}_{12}(\la_1,\la_2) M_2(\la_2) {\cal A}_{12}(\la_1,\la_2) 
\label{ecriture eqns compatibilite}
\enq
where 
\[
\wt{\cal B}_{12}^{\; t_1}(\la_1,\la_2) \; = \;   \Big({\cal B}_{12}^{\; t_1}(\la_1,\la_2)\Big)^{-1}   \qquad \e{and} \qquad 
\wt{\cal C}_{12}^{\; t_2}(\la_1,\la_2) \; = \;   \Big({\cal C}_{12}^{\; t_2}(\la_1,\la_2)\Big)^{-1} \;. 
\]
$\wt{M}_0(\la)$ being given, a generating function for conserved quantities is obtained as $\e{tr}_0\big[\op{T}_{0N}(\la) \wt{M}_0(\la)\big]$.

The most general solution to \eqref{ecriture eqns compatibilite} takes the form 
\beq
\wt{M}_0(\la) \; = \; \wt{\a} \left(\ba{cc} 1  &  \wt{\be} \la \\ 
				\wt{\ga} \la & q^{\frac{3}{2}}+ \wt{\de} \la + \wt{\be}  \la^2 \ea \right)_{[0]} \;, 
\label{matrice de cloture de la trace scalaire solution FM eqns}
\enq
for arbitrary constants $\wt{\a},\wt{\be},\wt{\ga},\wt{\de}$. 
The solution of interest to our analysis is obtained by taking 
\beq
\wt{\a}=q^{-1} \;,  \quad  \wt{\be} = q^{\frac{3}{2}} d_2 d_{3} \; , \quad \wt{\ga} = (1-q^{2}) \; ,  \quad  \wt{\de} = q^{\frac{5}{2}} d_{1} \;. 
\enq
  This specific solution is motivated by eq.(\ref{taut}) below.  It takes the factorised form  $\wt{G}_0(\la) q^{-\sg_0^z}$, where 
\begin{equation}
\wt{G}_0(\lambda) =  \left( \ba{cc} 1 & q^{-\frac{1}{2}} d_2 d_{3} \la  \vspace{2mm} \\ (1-q^2)\lambda &  q^{- \frac{1}{2} } + q^{\frac{1}{2} }d_{1} \lambda  + q^{-\frac{1}{2}} d_2 d_{3} \la^2   \ea \right) \;. 
 \label{gammamatrix}
\end{equation}
Thus, the generating function of conserved quantities is expressed as 
\beq
\bs{\tau}(\la) \; = \;  \;\e{tr}_0\big[\op{T}_{0N}(\la) \wt{G}_0(\la)q^{-\sg_0^z} \big] \;. 
\label{ecriture matrice transfert Toda2 non ultra loc}
\enq
These considerations show that the Toda$_2$ chain is a quantum integrable non-ultralocal  model that we could  already study
 at this level. However in the next section we will introduce quantum analogues of the Darboux coordinates eqs. \eqref{RepQn},\eqref{RepPn} which have the advantage 
 of ultra-localising the model.

 \section{Concrete realisation and ultra-localisation}
 \label{ultralocalisation}

 One can represent the quantum operators $\bs{\xi}_n$ in a way paralleling  the classical construction (\ref{DarbouxCoordinates}). 
One first introduces the local Hilbert spaces $\mf{h}_n\simeq L^2(\R)$ and constructs the full Hilbert space as  $\mf{h}=\otimes_{n=1}^{N} \mf{h}_n\simeq L^2(\R^N)$. 
Then, on each $\mf{h}_n$ one introduces a pair of canonically conjugate operators $\op{x}_n,\op{X}_n$ such that $[\op{x}_n,\op{X}_n]=\i$
and builds from them the Weyl pair
$$
\Big( \ex{\frac{2 \pi }{ \om_2} \op{x}_n }, \ex{\om_1 \op{X}_n }\Big) \qquad \e{so}\; \e{that} \qquad 
\ex{\frac{2 \pi }{ \om_2} \op{x}_n }  \ex{\om_1 \op{X}_n }  \; = \;  q^2 \,    \ex{\om_1 \op{X}_n }  \ex{\frac{2 \pi }{ \om_2} \op{x}_n } \;\; , \quad q= \ex{ \i \pi \f{\om_1}{\om_2} }\;. 
$$
Then,  
\beq
\bs{\xi}_n^1 \, = \,  \ex{ - \frac{ \pi }{ \om_2} \op{x}_n } \pl{a=1}{n}\ex{ \frac{\om_1}{2} \op{X}_n } \qquad \e{and} \qquad 
\bs{\xi}_n^1 \, = \,  \ex{ - \frac{ \pi }{ \om_2} \op{x}_n } \sul{a=1}{n} \pl{b=a}{n}  \ex{ \frac{\om_1}{2} \op{X}_b }   \cdot \ex{   \frac{ 2 \pi }{ \om_2} \op{x}_a } \cdot  \pl{b=1}{a-1}   \ex{ \frac{\om_1}{2} \op{X}_b }   
\enq
provides one with a representation on $\mf{h}$ of the algebra \eqref{ecriture algebre des xi}.

In terms of these canonical operators, one has
\begin{eqnarray*}
\op{Q}_n^2 &=& q^{1\over 2} \;  \ex{-\om_1 \op{X}_{n+1} }   \ex{\frac{2 \pi }{ \om_2} ( \op{x}_n - \op{x}_{n+1})} \; ,  \\
\op{P}_n &=&  \ex{-\om_1 \op{X}_{n } } \, + \,   \ex{\frac{2 \pi }{ \om_2} ( \op{x}_n - \op{x}_{n+1})} \;. 
\end{eqnarray*}
Written in these coordinates, the quantum Lax operator $\op{l}_{0n}(\lambda)$ is obviously not-ultralocal and reads 
$$
\op{l}_{0n}(\lambda) \; = \;  \left( \ba{cc}  \lambda - \ex{-\om_1 \op{X}_{n } } \, - \,   \ex{\frac{2 \pi }{ \om_2} ( \op{x}_n - \op{x}_{n+1})} & -1 \\
		q^{1\over 2}   \ex{-\om_1 \op{X}_{n+1} }   \ex{\frac{2 \pi }{ \om_2} ( \op{x}_n - \op{x}_{n+1})}  & 0 \ea \right) \;. 
$$

In the following, we will show that the generating function of conserved quantities given in \eqref{ecriture matrice transfert Toda2 non ultra loc}
can be recast in terms of a transfer matrix  built up from a monodromy matrix having an ultralocal structure. 

More precisely, define the ultralocal Lax matrix 
\beq
\op{L}_{0n} (\lambda)  \; = \;
\left(\ba{cc}   \la  - \ex{-\om_1 \op{X}_n }   &     q^{2} \la \big[ d_2 + q d_{1} \ex{-\om_1 \op{X}_n } + q^2 d_{3} \ex{- 2 \om_1 \op{X}_n } \big]  \ex{ - \frac{2 \pi }{ \om_2} \op{x}_n }   \\ 
	- q^{-2} \ex{   \frac{2 \pi }{ \om_2} \op{x}_n }  & -d_2  + \la d_{3} \ex{-  \om_1 \op{X}_n }    \ea \right) \;. 
  \label{Lultralocal}
\enq
This ultralocal Lax matrix satisfies the usual Yang-Baxter equation with quantum (twisted) $\op{R}$-matrix
\begin{equation}
\op{R}_{12}(\lambda_1,\lambda_2) = \left( \ba{cccc}1 & 0 & 0 & 0 \\
		    0 & {\lambda_2-\lambda_1 \over q^2\lambda_2-\lambda_1} & {(q^2-1) \lambda_1 \over q^2\lambda_2-\lambda_1}  & 0 \\
		    0 & {(q^2-1) \lambda_2 \over q^2\lambda_2-\lambda_1}  & {(\lambda_2-\lambda_1)q^2 \over q^2\lambda_2-\lambda_1}  & 0 \\
		    0 & 0 & 0 & 1 \ea \right) \;. 
\label{Rmatrix}
\end{equation} 
Let 
\beq
\op{t}_{\e{loc}}(\la) \; = \; \e{tr}\big[\op{L}_{0N} (\lambda) \cdots \op{L}_{01} (\lambda) \big]   \;,  
\label{definition matrice transfer locale}
\enq
be the associated transfer matrix and  let $\kappa_2$ parameterise $d_2$ as $d_2=\ex{-\frac{2\pi}{\om_2} \kappa_2}$. Then, it holds
\beq
d_2^{N}\cdot  q \cdot  \op{V} \cdot \bs{\tau}\big( q^{-2} d_2^{-1} \la \big) \cdot \op{V}^{-1} \; = \;\op{t}_{\e{loc}}(\la) \;. 
\label{taut}
\enq
where 
\beq
\op{V} \; = \; \pl{n=1}{N} \op{V}_n \qquad \e{and} \qquad \op{V}_n \, = \, \ex{ \frac{2\i \pi \kappa_2 }{ \om_1 \om_2}  \op{x}_n  } \cdot \ex{ \i \kappa_2 \op{X}_n } \;. 
\enq

In order to establish the result, first observe that one can transform the non-ultralocal monodromy matrix $\op{T}_{0N}(\la) \wt{G}_0(\la)q^{-\sg_0^z}$
by doing local gauge transformations:
\beq
\op{T}_{0N}(\la) \wt{G}_0(\la)q^{-\sg_0^z} \; = \; \op{N}_{0 N+1 } \;  \msc{L}_{0N}(\la) \cdots \msc{L}_{02}(\la) \wt{\msc{L}}_{01}(\la) \, \op{N}_{0  1 }^{-1} \, q^{-\sg_0^z}
\enq
where 
\[ 
\msc{L}_{0n}(\la)=\op{N}_{0 n+1 }^{-1} \, \wh{\op{l}}_{0n}( \la) \, \op{N}_{0 n }  \quad \e{for} \quad  n=2,\dots,N \quad  \e{and} \quad  \wt{\msc{L}}_{01}(\la) \; = \; \msc{L}_{01}(\la)_{\mid d_1\rightarrow q^{-2} d_1} \;. 
\]

The gauge matrices $ \op{N}_{0 k }$  are chosen as  
 $$
\op{N}_{0n} \;  = \; \left( \ba{cc} 1 & - q^{-\frac{1}{2}} \ex{ - \frac{2 \pi }{ \om_2} \op{x}_n }  \\ 0 & \ex{-\om_1 \op{X}_n } \ex{ - \frac{2 \pi }{ \om_2} \op{x}_n }  \ea \right)_{[0]} \; ,
\quad \e{so}\;\e{that} \quad 
\op{N}_{0n}^{-1} \;  = \; \left( \ba{cc} 1 &  q^{-\frac{1}{2}} \ex{ \om_1 \op{X}_n } \\ 0 &  \ex{   \frac{2 \pi }{ \om_2} \op{x}_n } \ex{ \om_1 \op{X}_n } \ea \right)_{[0]} \;. 
$$
The gauge transform of $\op{l}_{0n}(\lambda)$ already takes the ultralocal form 
$$
  \op{N}_{0 n+1}^{-1} \op{l}_{0n}(\lambda)\, \op{N}_{0n} \; = \; 
\left( \ba{cc} \lambda - \ex{-\om_1 \op{X}_n }  & \big[ - q^{-1/2} \lambda + (q^{-1/2}-1) \ex{-\om_1 \op{X}_n }  \big]  \ex{ - \frac{2 \pi }{ \om_2} \op{x}_n } \\  
	      q^{\frac{1}{2}}  \ex{   \frac{2 \pi }{ \om_2} \op{x}_n } &    - 1 						\ea \right) \;. 
$$
One still needs to deal with the matrix $G_0(\lambda)$:
\beq
G_{0n}(\la) \equiv  \op{N}_{0 n}^{-1}  G_0(\lambda)  \, \op{N}_{0n} \; = \;  \left( \ba{cc}  1 + (q^{-\frac{1}{2}}-q^{ \frac{3}{2} })\la \ex{\om_1 \op{X}_n }  
  &   \big[ G_{0n}(\la) \big]_{12}  \vspace{3mm} \\
(1-q^2)\la  \ex{   \frac{2 \pi }{ \om_2} \op{x}_n } \ex{\om_1 \op{X}_n }     &  \big[ G_{0n}(\la) \big]_{22} \ea \right) \; . 
\enq
Above, we agree upon 
\begin{align*}
\big[ G_{0n}(\la) \big]_{12}  & = \; 
\big[  q^{-1}+q^2d_{1}\la + q^{3} d_2 d_3 \la^2 - q^{-\frac{1}{2}}  + q^{\frac{7}{2}}d_2 d_3 \la \ex{- \om_1 \op{X}_n } -  (q^{-1}-q)\la \ex{\om_1 \op{X}_n }   \big]   \ex{ - \frac{2 \pi }{ \om_2} \op{x}_n }  \;, \\
 \big[ G_{0n}(\la) \big]_{22}  & = \;   q^{-\frac{1}{2}}+q^{\frac{5}{2}} d_{1}\la  + q^{\frac{7}{2}}d_2 d_3 \la^2 - \la (q^{\frac{3}{2}}-q^{\frac{7}{2}})\ex{\om_1 \op{X}_n }   \;. 
\end{align*}
Hence, all-in-all , one obtains 
\beq
\msc{L}_{0n}(\la) \; = \;
\left(\ba{cc}   q^2\la  - \ex{-\om_1 \op{X}_n }   &  - q^{\frac{3}{2}} \la \big[ 1 + q d_{1} \ex{-\om_1 \op{X}_n } +q^2 d_{2} d_3 \ex{-2\om_1 \op{X}_n } \big]  \ex{ - \frac{2 \pi }{ \om_2} \op{x}_n }   \\ 
	q^{\frac{1}{2}} \ex{   \frac{2 \pi }{ \om_2} \op{x}_n }  & -1   + q^2 \la d_{2} d_3 \ex{- \om_1 \op{X}_n }    \ea \right) \;. 
\enq
Then straightforward algebra and the use of periodic boundary conditions $\op{N}_{0 N+1 } = \op{N}_{0 1}$ leads to 
\beq
\e{Tr}_{0}\Big[\op{N}_{0 N+1 } \msc{L}_{0N}(\la) \cdots \msc{L}_{02}(\la) \wt{\msc{L}}_{01}(\la)\op{N}_{0  1 }^{-1}  q^{-\sg_0^z}  \Big] \; = \; 
q^{-1} \cdot \e{Tr}_{0}\Big[  \msc{L}_{0N}(\la) \cdots \msc{L}_{01}(\la)  \Big] \;. 
\enq
The equality can be obtained by first taking explicitly the matrix products on the auxiliary space of  $\op{N}_{01 }^{-1}$, $ \msc{L}_{0N}(\la) \cdots \msc{L}_{02}(\la)$ and $ \wt{\msc{L}}_{01}(\la)\op{N}_{0  1 } q^{-\sg_0^z}$
and then by taking the trace. Finally one compares this with the result of a similar calculation carried out on the level of the \textit{rhs} expression.  

It now only remains to observe that it holds 
\beq
d_2 \, \sg_0^z \, q^{\frac{5}{4}\sg^{z}_0} \, \op{V}_n \cdot\msc{L}_{0n}\big( q^{-2} d_2^{-1} \la \big) \cdot   \op{V}_n^{-1} \,  \sg_0^z \,  q^{-\frac{5}{4} \sg^{z}_0 } \; = \;\op{L}_{0n}(\la) \; . 
\enq

 {\bf Local Hamiltonians.}
 Upon expanding the transfer matrix into powers of $\la$:
\beq
\op{t}_{\e{loc}}(\la) \; = \; \e{tr}\big[\op{L}_{0N} (\lambda) \cdots \op{L}_{01} (\lambda) \big] \; = \;   \sul{j=0}{N} (-1)^j\cdot \la^{N-j}  \cdot  \op{H}_j   \;,  
\enq
one gets a  family of commuting Hamiltonians $\{ \op{H}_1, \dots,  \op{H}_N \}$.
There are three interesting limiting cases.   

$\bullet$ {\bf $q$-Toda.}
For $d_2=d_3=0$, one gets the $q$-Toda chain 
\[
\op{H}_1^{\mathrm{q-Toda}}= \sum\limits_{n=1}^N \left[ 1+ q^{-1} d_1 \ex{-{2\pi \over \omega_2}(\op{x}_n-\op{x}_{n-1})}  \right]\ex{-\omega_1 \op{X}_n} 
\]

 $\bullet$ {\bf Toda$_2$.}
For $d_1=d_3=0$, one gets the \textit{per se} Toda$_2$ chain. We find
\[
\op{H}_1^{\mathrm{Toda}_2} =  \sum\limits_{n=1}^N  \Big\{ \ex{-\omega_1 \op{X}_n} + d_2\ex{-{2\pi \over \omega_2}(\op{x}_n-\op{x}_{n-1})}  \Big\}\;. 
\]
\begin{empheq}{align*}
\op{H}_2^{\mathrm{Toda}_2}-{1\over 2} (\op{H}_1^{\mathrm{Toda}_2})^2 &= -{1\over 2}  \sum\limits_{n=1}^N \Big\{ \ex{-2\omega_1 \op{X}_n}   \\
& \hskip-2cm +  d_2  \ex{-\omega_1 \op{X}_n} \left[(1+q^{-2}) \ex{{2\pi \over \omega_2} ( \op{x}_n- \op{x}_{n+1})} + (1+q^2) \ex{ {2\pi \over \omega_2} ( \op{x}_{n-1} - \op{x}_{n})}  \right] 
+ d_2^2 \, \ex{ {4\pi \over \omega_2} ( \op{x}_n - \op{x}_{n+1} ) } \Big\} \;. 
\end{empheq}
In the absence of a quantum analogue of eq.(\ref{poissonL}) and a suitable definition of ${\rm tr}_q$, it is not possible to define the  family of quantum Hamiltonians 
of Toda$_2$ directly in terms of the  big quantum Lax matrix $L(\mu)$. However for $\op{H}_1^{\mathrm{Toda}_2}$  we obviously have
$$
\op{H}_1^{\mathrm{Toda}_2}\vert_{d_2=1} \equiv - {\rm tr}_q\big[ L(\mu) \big] \,  =  \, \sum_{n=1}^N \op{P}_n \;. 
$$
We easily find  the correct definition of ${\rm tr}_q [L^2(\mu)]$  by imposing the commutativity with ${\rm tr}_q\big[ L(\mu) \big]$
and the fact that in the classical limit $q \rightarrow 1$, ${\rm tr}_q [L^2(\mu)]$ should reduce to ${\rm tr} [L^2(\mu)]$, 
$$
{\rm tr}_q [L^2(\mu)] \equiv \sum_{n=1}^{N} \Big\{ \op{P}_n^2 + (q^{3/2}+q^{-1/2}) \op{Q}_n^2 \Big\}
$$
Inserting the ultralocal parameterisation of $\op{P}_n,\op{Q}_n^2$,  we discover that
$$
\left[ \op{H}_2^{\mathrm{Toda}_2}-{1\over 2} (\op{H}_1^{\mathrm{Toda}_2})^2 \right]_{  \mid d_2=1} = -{1\over 2} {\rm tr}_q [L^2(\mu)] 
$$
These formulae support the evidence that the quantum Toda$_2$ model is indeed obtained by the choice of parameters
$d_1=0, d_2=1, d_3=0$ in the above family of models.

\vspace{2mm}

 $\bullet$ {\bf $q$-oscillator.} When $d_1=-q^{-1},d_2=1$ and $d_3=0$, the local Lax  operator given above can also be rewritten in terms of $q$-oscillator.
  
  Let
\begin{empheq}{align*}
\op{a}_n = (1- \ex{-\omega_1 \op{X}_n}) \ex{-{2\pi \over \omega_2} \op{x}_n} \, ,\quad 
\op{a}_n^*  = \ex{{2\pi \over \omega_2} \op{x}_n}, \quad  \ex{-\omega_1 \op{X}_n}  \, = \,  q^{2\op{D}_n}
\end{empheq}
then the $q$-oscillators algebra is satisfied
\begin{empheq}{align*}
\op{a}_n \op{a}_n^* &= 1 - q^{2\op{D}_n}, \quad \op{a}_n^* \op{a}_n  = 1 -  q^{2\op{D}_n-2}\\
\op{a}_n q^{2\op{D}_n } &= q^{2\op{D}_n+2}\op{a}_n, \quad \op{a}_n^* q^{2\op{D}_n } = q^{2\op{D}_n-2} \op{a}_n^* 
\end{empheq}
The local Lax  operator then becomes
\begin{equation}
\op{L}_{0n} (\lambda)  \; = \;
\left( \ba{cc}   \la  - q^{2\op{D}_n}  &     q^{2} \la\;   \op{a}_n   \\ 
	- q^{-2} \op{a}_n^*  & -1   \ea \right) \;. 
\label{ecriture matrice de Lax q-oscillateurs}
\end{equation}

Choose a complete set of left Eigenstates associated with the $q$-boson oscillator algebra $\big\{ \bs{v}_{n}^{(k)} \big\}_{k \geq 0}$, \textit{viz}. 
\beq
\bs{v}_{n}^{(k)} \op{a}_n \, = \, \big( 1 - q^{-2k} \big) \bs{v}_{n}^{(k-1)} \; , \quad \bs{v}_{n}^{(k)} \op{a}_n^* \, = \,  \bs{v}_{n}^{(k+1)} 
\; , \quad \bs{v}_{n}^{(k)} q^{2\op{D}_n } \, = \, q^{ - 2k } \bs{v}_{n}^{(k)} \;. 
\enq
Next consider $\bs{\om}_n$ to be the left Eigenstate of $\op{a}_n$ with eigenvalue $q^{-2}$:
\beq
\bs{\om}_n  \op{a}_n\, = \, q^{-2} \bs{\om}_n \qquad \e{so}\, \e{that} \qquad 
\bs{\om}_n \,  =  \, \sul{k=0}{+\infty}   {q^{-2k}\over (1-q^{-2})\cdots(1-q^{-2k}) } \bs{v}_{n}^{(k)} \;. 
\label{defomega} 
\enq
$\bs{\om}_n$ is an Eigenstate of the sum of the matrix elements of each column of the q-oscillator Lax matrix \eqref{ecriture matrice de Lax q-oscillateurs}  with
eigenvalue $\lambda-1$. This property readily follows from the identity: $-q^{-2}\bs{\om}_n  \op{a}_n^* \, = \, \bs{\om}_n \big( q^{2\op{D}_n} -1 \big)$. Consequently, $ \bs{\Om}=\otimes_{n=1}^N \bs{\om}_n $ is a left Eigenstate of the Hamiltonian
\begin{eqnarray}
\op{H}_1 \, = \, \sul{n=1}{N} \Big\{  \op{a}_n \op{a}^*_{n+1}+q^{2D_n}  \Big\}
\label{ham} 
\end{eqnarray}
with eigenvalue $N$. So, if we define a scalar product on each local space such that $\big( \bs{\om}_n,\bs{v}_{n}^{(k)} \big) \,= \, 1 \, , \;  \forall k$,  
then the Hamiltonian $\op{H}\, - \, N\e{id}$ defines a stochastic model \cite{SW98}.

\section*{Conclusion}

In the present work we have constructed the appropriate quantisation of the Toda chain endowed with the second Hamiltonian structure. 
This model appears as a natural lattice discretisation of the quantum Viasoro algebra and may appear in the future as a convenient way to 
build explicit representations of this algebra, in the continuum, for arbitrary values of the central charge. 
Our construction builds on the quantum ultra-localisation of the non-ultra local algebra appearing naturally in the context
of the Toda chain in the second Hamiltonian formulation.

\section*{Acknowledgment}
O.B. thanks E. K. Sklyanin for stimulating discussions.


\begin{thebibliography}{XXXXX}
 
 \bibitem{Adl79} M. Adler. {\it ``On a trace functional for formal pseudodifferential operators and symplectic structure of the Korteweg-de Vries
type equations''}. Inv. Math. 50 (1979), p. 219.

 




\bibitem{Ba88} O.Babelon,  {\it ``Extended Conformal Algebra and the  Yang-Baxter  Equation"} Physics  Letters 215B (1988) 523-529.

 \bibitem{Ba90} O. Babelon, {\it Exchange formula and lattice deformation of the Virasoro algebra.} Phys.Lett. B238 (1990) 234. 


\bibitem{Dam94} P. Damianou {\it Multiple Hamiltonian Structures for Toda-type systems.} Journal of Mathematical Physics, 35 5511-5541 (1994). 

 \bibitem{GN84} J.L.Gervais, A.Neveu, {\it Novel triangle relation and absence of tachyon in Liouville string field theory.} Nucl.Phys. B238 (1984) 125.
 
 
\bibitem{FaTa86} L.D.Faddeev, L.Takhtajan,  {\it Liouville model on the lattice.} Springer Lectures notes in Physics, 246 (1986) 66.




%
%
%
%

%
%
%

\bibitem{FreMa91a} L. Freidel, J-M. Maillet, {\it Quadratic algebras and integrable systems.} Phys. Lett. 262B (1991) 278-284.

\bibitem{FreMa91b} L. Freidel, J-M. Maillet, {\it On the classical and quantum integrable field theories associated to Kac-Moody current algebras.} 
Phys. Lett. 263B (1991) 403-410.





%
%
%
%
%
%
%
%
%


\bibitem{SW98} T. Sasamoto and M. Wadati, {\it Exact results for one-dimensional totally asymmetric diffusion models}, J. Phys. A 31 (1998), 60576071.

\bibitem{V92} A. Volkov, {\it Quantum Volterra model.} Phys. Letters A, 167, (1992) 345-355.

\end{thebibliography}
\end{document}